\documentclass[prd,aps,floats,preprint,preprintnumbers,amssymb,nofootinbib]{revtex4}
\usepackage{amsmath,amssymb}
\usepackage{graphicx}
\usepackage{graphics}
\usepackage{epsfig}
\usepackage{latexsym,oldgerm}

\def\beq{\begin{equation}}
\def\eeq{\end{equation}}
\def\be{\begin{equation}}
\def\ee{\end{equation}}

\def\bea{\begin{eqnarray}}
\def\eea{\end{eqnarray}}

\begin{document}
\title{Non-Pauli Transitions From Spacetime Noncommutativity}
\author{A. P. Balachandran$^{a,b}$}\thanks{bal@phy.syr.edu}
\author{Anosh Joseph$^{a}$}\thanks{ajoseph@phy.syr.edu}
\author{Pramod Padmanabhan$^{a,b}$}\thanks{ppadmana@syr.edu}
\affiliation{$^{a}$Department of Physics, Syracuse University,
Syracuse, NY 13244-1130, USA}\affiliation{$^{b}$ The Institute of
Mathematical Sciences, CIT Campus, Taramani, Chennai, India 600 113}

\begin{abstract}
There are good reasons to suspect that spacetime at Planck scales is
noncommutative. Typically this noncommutativity is controlled by
fixed ``vectors'' or ``tensors'' with numerical entries. For the
Moyal spacetime, it is the antisymmetric matrix $\theta_{\mu\nu}$.
In approaches enforcing Poincar\'e invariance, these deform or twist
the method of (anti-)symmetrization of identical particle state
vectors. We argue that the earth's rotation and movements in the
cosmos are ``sudden'' events to Pauli-forbidden processes. They
induce (twisted) bosonic components in state vectors of identical
spinorial particles in the presence of a twist. These components
induce non-Pauli transitions. From known limits on such transitions,
we infer that the energy scale for noncommutativity is $\gtrsim
10^{24}~\textrm{TeV}$. This suggests a new energy scale beyond
Planck scale.
\end{abstract}

\date{\today}
\maketitle
\preprint{SU-4252-901}

\section{Introduction}
The consideration of noncommutative spacetimes in quantum theory can
be plausibly advocated from physics at the Planck
scale~\cite{Doplicher:1994zv}. A popular spacetime for this purpose
has been the Moyal plane $\mathcal{A}_\theta$, a deformed spacetime
algebra~\cite{Akofor:2008ae}. Here instead we consider another
algebra $\mathcal{B}_{\chi\hat{n}}$ which is better suited to study
atomic processes. If $\hat{x}_\mu$, are coordinate functions on
spacetime, \beq \hat{x}_\mu(x) = x_\mu, \eeq then this algebra is
defined by the relation \beq \label{Bplane} \left[\hat{x}_0,
\hat{x}_j\right] = i\chi \epsilon_{ijk} n_i\hat{x}_k, \eeq where
$i\in [1,2,3]$, $\chi \in \mathbb{R}$ and $\vec{n}$ is a unit
vector. Here $\chi$ and $\vec{n}$ are fixed and not dynamical. We
can also introduce noncommutativity between components of
$\vec{n}\wedge\vec{x}$, but we will not do so.

The Poincar\'e group $\mathcal{P}$ can be implemented on
$\mathcal{B}_{\chi\hat{n}}$ provided the canonical coproduct
$\Delta_0$ of the group algebra $\mathbb{C}\mathcal{P}$, \beq
\Delta_0(g) = g\otimes g,~~ g\in\mathcal{P} \eeq is deformed to
$\Delta_{\chi\hat{n}}$ by a Drinfel'd twist
$\mathcal{G}_{\chi\hat{n}}$: \beq \Delta_{\chi\hat{n}}(g) =
\mathcal{G}_{\chi\hat{n}}^{-1}\Delta_0(g)\mathcal{G}_{\chi\hat{n}}
\eeq \beq \mathcal{G}_{\chi\hat{n}} =
e^{-i\frac{\chi}{2}\left(P_0\otimes\hat{n}\cdot \vec{J} -
\hat{n}\cdot\vec{J}\otimes P_0\right)}. \eeq For reviews
see~\cite{Akofor:2008ae}. Here $P_0$ and $\vec{J}$ are time
translation and rotation generators of $\mathcal{P}$.

Let $\mathcal{H}$ be a representation space of
$\mathbb{C}\mathcal{P}$. Then we can define the flip operator
$\tau_0$ on $\mathcal{H}\otimes\mathcal{H}$: \beq \tau_0 (v\otimes
w) = (w \otimes v),~~~v,w~\in~ \mathcal{H}.\eeq Since $\tau_0$
commutes with the action of $\Delta_0(g)$ on
$\mathcal{H}\otimes\mathcal{H}$, we can symmetrize and
anti-symmetrize $\mathcal{H}\otimes\mathcal{H}$ for $\chi=0$ using
the projectors $\frac{1}{2}(1\pm\tau_0)$ to get untwisted bosons and
fermions. But $\tau_0$ does not commute with the action of
$\Delta_{\chi\hat{n}}$ if $\chi\neq 0$. Instead the twisted
symmetrizer \beq \tau_{\chi\hat{n}} =
\mathcal{G}_{\chi\hat{n}}^{-1}\tau_0\mathcal{G}_{\chi\hat{n}}=
\mathcal{G}_{\chi\hat{n}}^{-2}\tau_0, \eeq \beq
\mathcal{G}_{\chi\hat{n}}^{-2} =
e^{i\chi\left(P_0\otimes\hat{n}\cdot\vec{J} -
\hat{n}\cdot\vec{J}\otimes P_0\right)} \eeq does. The projectors
$\frac{1}{2}(1\pm\tau_{\chi\hat{n}})$ on
$\mathcal{H}\otimes\mathcal{H}$ then give the twisted bosons and
fermions.

Such twisted (anti-)symmetrization can be extended to
$\mathcal{H}^{\otimes k}$ for higher values of $k$.

Twisted anti-symmetrization induces $\chi\hat{n}$ dependence in
energy eigenstates of electrons (nucleons) in atoms (nuclei) and
corrects lifetimes in atomic (nuclear) processes. These corrections
are expected to have very long time scales, $\chi$ being of the
order of Planck length. (The corrections to rates from $\chi$ are
$O(\chi^2 E)$ and the corresponding times are $O(\chi^{-2}E^{-1})$.
$E$ is the typical energy involved in such transitions. Note also
that we do not remark on TeV gravity till Sec.IV.) They are expected
to be much longer than terrestrial times like 24 hours or 1 year.
Thus the earth's motions are sudden for noncommutative effects.
 But the earth for example is a rotating frame and not an
inertial frame. $J_i$ changes in that frame to $R_{ik}(t)J_k$ where
we can permit $R(t)\in SO(3)$ to have dependence on time $t$. That
changes $\hat{n}\cdot\vec{J}$ to $\hat{m}\cdot\vec{J}$, $m_i := n_k
R_{ki}(t)$ and hence the twisted flip operator to
$\tau_{\chi\hat{m}}$. Thus {\it effectively}, the non-dynamical
$\vec{n}$ gets rotated to $\vec{m}$.  An effect of this sort has
been noticed before by~\cite{Piacitelli:2009fa}. The swift change of
$\tau_{\chi\hat{n}}$ to $\tau_{\chi\hat{m}}$ induces (twisted)
bosonic components in multi-fermion state vectors and leads to
non-Pauli effects.

 We can express these effects in another manner.
The energy eigenstates for the twisted flip $\tau_{\chi\hat{n}}$
depend on $\hat{n}$. When $\hat{n}$ suddenly changes to $\hat{m}$
due to the earth's motions, these states do not change in the sudden
approximation. But they are not eigenstates for the flip
$\tau_{\chi\hat{m}}$. When expanded in $\tau_{\chi\hat{m}}$
eigenstates, they are found to have $\tau_{\chi\hat{m}} = 1$
(twisted boson) components as well. These cause the non-Pauli
effects.

Such non-Pauli effects are expected for the Moyal twist $\tau_\theta$
as well~\cite{Balachandran:2005eb}.

There is a trinity of time scales, atomic($\tau_-$),
terrestrial($\tau_0$) and time for a Pauli-forbidden transition to
occur($\tau_+$), fulfilling $\tau_-<<\tau_0<<\tau_+$ where
``atomic'' refers to ``nuclear'' as well. The earth's movements are
extremely adiabatic processes for atomic dynamics and are not
important (unless through Berry phase effects). But that is not the
case for physics with typical times $\tau_+$ for which the earth's
motions are ``sudden''.

\section{The Details}
We focus on the neutral Be atom with its $4$ electrons for
specificity. Let $\vec{X}^{(\alpha)}$ ($\alpha = 1, 2$) be the
coordinate functions of the electrons in $Be$ and $\vec{X}$ that of
the nucleus. (We drop the hat on $\hat{X}$'s). Each of them, and
hence also their differences; fulfill Eq. (\ref{Bplane}). In
particular the relative coordinates \beq \vec{x}^{(\alpha)} =
\vec{X}^{(\alpha)} - \vec{X} \eeq fulfill \beq \left[x_0,
x_j^{(\alpha)}\right] = i\chi\epsilon_{ijk}n_i x_k^{(\alpha)}. \eeq
For the Moyal plane, the relative coordinates and $x_0$ all mutually
commute~\cite{Balachandran:2004cr} forcing us to consider
relativistic kinematics where center-of-mass and relative motions
influence each other. That is why we consider
$\mathcal{B}_{\chi\hat{n}}$.

\subsection{\it{Preliminaries}}
Let $P_0$ be the single electron Hamiltonian
$-\frac{\vec{\nabla}^{2(\alpha)}}{2\mu} -
\frac{4e^2}{|\vec{x}^{(\alpha)}|}$, where $\mu$ is the reduced mass.
It represents $i\partial_t$ on single electron wave functions. On
two-electron states, it acts as \beq \Delta_{\chi\hat{n}}(P_0) =
P_0\otimes\mathbb{I} + \mathbb{I}\otimes P_0, \eeq $P_0$ commuting
with $J_i$. As for coproducts of angular momentum $\vec{J}$, let
$\hat{n}$, $\hat{n}^{(1)}$ and $\hat{n}^{(2)}$ form an orthonormal
frame with $\hat{n}^{(1)}\wedge\hat{n}^{(2)} = \hat{n}$ and let
$\vec{n}^{(\pm)} = \hat{n}^{(1)} \pm i\hat{n}^{(2)}$. Then using
$\left[\hat{n}\cdot\vec{J}, \vec{n}^{(\pm)}\cdot\vec{J}\right] = \pm
\vec{n}^{(\pm)}\cdot\vec{J}$, we find \beq \label{n.j}
\Delta_{\chi\hat{n}}(\hat{n}\cdot\vec{J})=\hat{n}\cdot\vec{J}\otimes
\mathbb{I} + \mathbb{I}\otimes \hat{n}\cdot\vec{J}, \eeq \beq
\label{npm.j} \Delta_{\chi\hat{n}}(\vec{n}^{(\pm)}\cdot\vec{J}) =
\vec{n}^{(\pm)}\cdot\vec{J}\otimes e^{\mp i\frac{\chi}{2}P_0} +
e^{\pm i\frac{\chi}{2}P_0}\otimes\vec{n}^{(\pm)}\cdot\vec{J}. \eeq

Our basic Pauli-forbidden process is that of two electrons in an
excited state transiting to the ground two-electron state already
occupied by two electrons. This transition can be caused by a
generic perturbation $V_{\chi\vec{n}}$ of the two-electron
Hamiltonian. For $\chi=0$, for simplicity, we take $V_0$ to be
spin-independent, like the Coulomb repulsion
$\frac{e^2}{|\vec{x}^{(1)}-\vec{x}^{(2)}|}$, between the two
electrons. As $V_0$ must commute with $\tau_0$, it is also symmetric
in the electron coordinates. In the presence of the twist, the
perturbation, just as $\Delta_{\chi\vec{n}}(P_0)$ and all
observables must commute with $\tau_{\chi\vec{n}}$, making us modify
$V_0$ to \beq V_{\chi\vec{n}} = \frac{1}{2}\left[V_0 +
\tau_{\chi\vec{n}}V_0\tau_{\chi\vec{n}}\right]. \eeq So the
two-electron Hamiltonian \beq H^{(2)} = \Delta_{\chi\vec{n}}(P_0) +
V_{\chi\vec{n}} \eeq is $\chi\vec{n}$ dependent.

{\it Remark:} Hopf symmetry, like any other symmetry can be broken.
Since $H^{(2)} \neq \Delta_{\chi\vec{n}}(P_0)$, the Hopf symmetry
generated by $P_0$ and $\vec{J}$ is broken.

We consider only orbital angular momentum $l=0$ energy levels for
ease of calculation, and choose a basis of spin states $|\alpha
\rangle_{\vec{n}}$ ($\alpha=\pm 1, \textrm{often denoted as
just}~\pm $) polarized in direction $\vec{n}$: \beq
\frac{\vec{\sigma}\cdot\vec{n}}{2}|\alpha\rangle_{\vec{n}} =
\frac{\alpha}{2}|\alpha\rangle_{\vec{n}},~~~\sigma_i = \textrm{Pauli
matrices}. \eeq Then if $|\nu\rangle$ are the radial single electron
states for principal quantum number $\nu$, we write \beq |\nu
\rangle\otimes |\alpha\rangle_{\vec{n}} = |\nu ,
\alpha\rangle_{\vec{n}}, \eeq \beq |\nu , \alpha \rangle_{\vec{n}}
\otimes |\nu ' , \beta \rangle_{\vec{n}} = |\nu , \alpha ; \nu ' ,
\beta \rangle_{\vec{n}}. \eeq The energy of $|\nu ,
\alpha\rangle_{\vec{n}}$ is called $E_{\nu}$: \beq P_0 |\nu , \alpha
\rangle_{\vec{n}} = E_{\nu}|\nu , \alpha \rangle_{\vec{n}} \eeq
while that of $|\nu , \alpha ; \nu ' , \beta \rangle_{\vec{n}}$, on
ignoring $V_{\chi\vec{n}}$ is $E_{\nu} + E_{\nu '}$: \beq
\Delta_{\chi\vec{n}}(P_0) |\nu , \alpha ; \nu ' ,
\beta\rangle_{\vec{n}} = (E_{\nu}+E_{\nu '})|\nu , \alpha ; \nu ' ,
\beta \rangle_{\vec{n}}. \eeq

\subsection{\it{The Ground State and Excited States}}
The normalized twist-antisymmetrized two-electron ground state is
\beq
\label{ground}
\begin{split}|1,1\rangle_{\chi\vec{n}} &= \sqrt{2}\frac{1-\tau_{\chi\vec{n}}}{2}\left[|1+, 1 - \rangle_{\vec{n}} \right] \\
&= \frac{1}{\sqrt{2}}\left[|1+,1-\rangle_{\vec{n}} - e^{i\chi
E_1}|1-,1+\rangle_{\vec{n}}\right] \\ &= -e^{i\chi
E_1}\sqrt{2}\frac{1-\tau_{\chi\vec{n}}}{2}(|1-,1+\rangle_{\vec{n}}).\end{split}
\eeq
\beq
\Delta_{\chi\vec{n}}(P_0)|1,1\rangle_{\chi\vec{n}} = 2 E_1 |1, 1 \rangle_{\chi\vec{n}}
\eeq

For $\chi=0$, the ground state, a spin singlet, is unique. By
continuity, it remains so for $\chi \neq 0$. For this reason,
replacement of either $|\pm\rangle_{\vec{n}}$ in Eq. (\ref{ground})
by other spin states does not give new answers.

We put two of the electrons in the above ground state.

We put the remaining two electrons in the $s$-wave levels with
$\nu = 2 ~\textrm{and}~ 3$. Consider for specificity their state
\beq
\begin{split}
|2+,3+\rangle_{\chi\vec{n}}  & =
\frac{1-\tau_{\chi\vec{n}}}{\sqrt{2}}|2+,3+\rangle_{\vec{n}} \\ & =
\frac{1}{\sqrt{2}}\left[|2+,3+\rangle_{\vec{n}} -
e^{i\frac{\chi}{2}(E_3-E_2)}|3+,2+\rangle_{\vec{n}}\right]
\end{split}
\eeq When the world turns, the Hamiltonian becomes
$\Delta_{\chi\vec{m}}(P_0) + V_{\chi\vec{m}}$. The projectors to its
anti-symmetrized eigenstates, in particular, the projector,
$|1~1\rangle_{\chi\vec{m}~\chi\vec{m}}\langle 1~1|$ is an observable
because the Hamiltonian is an observable. So in particular the
Hilbert space of states contains
$\mathbb{C}|1~1\rangle_{\chi\vec{m}}$.

But in the sudden approximation, it contains
$|1~1\rangle_{\chi\vec{n}}$ as well. We now show that it is not
orthogonal to the $\tau_{\chi\vec{m}} = +1$ state \beq
\frac{1+\tau_{\chi\vec{m}}}{2}|1+,1+\rangle_{\vec{m}} =
|1+,1+\rangle_{\vec{m}} \eeq This follows from~\cite{Future}, \beq
|_{\vec{m}}\langle \rho |\alpha\rangle_{\vec{n}}|^2 =
\frac{1}{2}\left[1+(-1)^{\frac{(\rho-\alpha)}{2}}\vec{m}\cdot\vec{n}\right]
\eeq so that \beq
|_{\vec{m}}\langle1+,1+|1~1\rangle_{\chi\vec{n}}|^2 =
\frac{1}{2}\left[1-(\vec{m}\cdot\vec{n})^2\right]\sin^2(\frac{\chi
E_1}{2}) \neq 0 \eeq if $\vec{m}\neq\vec{n}$, $\chi\neq 0$.

Thus $|1~1\rangle_{\chi\vec{n}}$, which is in the Hilbert space of
states, is linearly independent of the $\tau_{\chi\vec{m}} = -1$
vector $|1~1\rangle_{\chi\vec{m}}$. Hence the Hilbert space contains
at least one vector with energy $2E_1$ perpendicular to
$|1~1\rangle_{\chi\vec{m}}$, namely  $|1~1\rangle_{\chi\vec{n}}
-_{\chi\vec{m}} \langle
1~1|1~1\rangle_{\chi\vec{n}}|1~1\rangle_{\chi\vec{m}}$. It is part
of a spin triplet. But $\Delta_{\chi\vec{m}}(J_i)$ are observables,
form an angular momentum algebra, and its triplet representation is
irreducible. So now the ground state is enhanced to contain the
entire triplet of angular momentum one states. The projector \beq
\label{Projector} P =
\mathbb{I}_{2E_1}-|1,~1\rangle_{\chi\vec{m}~\chi\vec{m}}\langle
1,~1|, \eeq \beq \mathbb{I}_{2E_1} = |1,~1\rangle\langle 1,~1|, \eeq
\beq |1,~1\rangle := |\nu =1\rangle\otimes |\nu =1\rangle \eeq to
the subspace of these states is also an observable.

We can check that $|2+,3+\rangle_{\chi\vec{n}}$ contains bosonic components
for the twist $\tau_{\chi\vec{m}}$:
\beq
\frac{1+\tau_{\chi\vec{m}}}{2}|2+,3+\rangle_{\chi\vec{n}}\neq 0.
\eeq

We now calculate the rate for the transition \beq
|2+,3+\rangle_{\chi\vec{n}}  \rightarrow
\frac{1+\tau_{\chi\vec{m}}}{\sqrt{2}}|1\alpha
,1\beta\rangle_{\vec{m}} \eeq due to the potential
$V_{\chi\vec{m}}$. We can neglect its $\chi$ dependence which only
introduces $O(\chi^2)$ corrections in the transition amplitude. The
perturbation $V_0$ is symmetric in electron coordinates as it must
commute with $\tau_0$. By assumption, it is spin-independent.

Then if at time $t_i$, the two electrons were in the
$|2+,3+\rangle_{\chi\vec{n}}$ state, the transition probability
$\mathcal{P}(t_f,t_i)$ to any of the three bosonic ground state at
time $t_f$ is \beq
\begin{split}
\mathcal{P}(t_f,t_i) &=
_{\chi\vec{n}}\langle 2+,3+|\int_{t_i}^{t_f}d\tau 'e^{-iH_0\tau
'}V_0(\tau ')e^{i H_0\tau '}P \int_{t_i}^{t_f}d\tau e^{-iH_0\tau
}V_0(\tau)e^{i H_0\tau}|2+,3+\rangle_{\chi\vec{n}} \\ & =|\langle
1,~1|\int_{t_i}^{t_f} d\tau e^{i\tau
2E_1}V_0(\tau)e^{-i\tau(E_2+E_3)}|2,~3\rangle|^2\times
P_{\textrm{SPIN}}^{\chi}
\end{split}
\eeq where $P$ is in Eq. (\ref{Projector}), $\langle
1,~1|V_0|2,~3\rangle$ is the radial matrix element of $V_0$ and \beq
\begin{split}
P_{\textrm{SPIN}}^{\chi} &=
\frac{1}{2}|(1-e^{\frac{i\chi}{2}(E_3-E_2)})|^2\left[1-\frac{1}{2}|(_{\vec{m}}\langle
+-|-e^{-i\chi E_1}~_{\vec{m}}\langle
-+|)|++\rangle_{\vec{n}}|^2\right] \\ &=
\frac{1}{2}|(1-e^{i\frac{\chi}{2}(E_3-E_2)})|^2\left[1-\frac{1}{2}|1-e^{-i\chi
E_1}|^2\frac{1}{4}(1-(\vec{m}\cdot\vec{n})^2)\right]
\end{split}
\eeq

Since $\vec{m}$ and $\vec{n}$ keep changing, now average
$\mathcal{P}(t_f,t_i)$ over the directions of $\vec{m}$ and
$\vec{n}$ using the standard rotationally invariant measure $d\Omega
= \frac{1}{4\pi}d\cos(\theta)d\cos\phi$. Then \beq \langle
m_i\rangle\equiv \int \frac{d\Omega_{\vec{m}}}{4\pi} m_i = 0, \eeq
\beq \langle m_i m_j \rangle \equiv \int
\frac{d\Omega_{\vec{m}}}{4\pi} m_i m_j = \frac{1}{3} \delta_{ij}
\eeq and \beq \label{rate} \langle \mathcal{P}(t_f,t_i)\rangle =
\left[|\langle 1,~1|\int_{t_i}^{t_f} d \tau e^{i\tau
2E_1}V_0(\tau)e^{-i\tau(E_2+E_3)}|2,~3\rangle|^2\right]\times\left[\frac{1}{3}(5+\cos(\chi
E_1))\sin^2(\frac{\chi}{4}\Delta E)\right], \eeq \beq \Delta E = E_3
- E_2 . \eeq

Similar probabilities can be deduced for different initial and final
states. The answers will not be Eq. (\ref{rate}), but they are still
expected to be $O(\chi ^2)$. Our bounds for $\chi$ below are not
expected to change much by such changes.

It is best to work with the branching ratio $B_{\chi}$ of the
Pauli-forbidden process to an allowed process to cancel out the
details specific to our model and give a formula of general
applicability. The terms multiplying the $\chi$ dependent part is
expected to approximate a typical Pauli-allowed process. Thus the
branching ratio of a Pauli-forbidden to an allowed process is \beq
\label{ratio} B_{\chi} = \frac{1}{3}\left[5+\cos(\chi
E_1)\right]\sin^2(\frac{\chi}{4}\Delta E). \eeq

\section{The Bounds}
We can now use $B_{\chi}$ for both atomic and nuclear
experiments~\cite{Borexino, NEMO, NEMO-2, Ramberg, SKamiokande, VIP}
to deduce bounds on $\chi$, with $\Delta E$ standing for the change
in energy in the nuclear or atomic transition. It also indicates
whether the experiment involves nuclear or atomic transition.

The bounds are summarized in Table (\ref{table1}). They are obtained
from the following experimental branching ratios:

Borexino~\cite{Borexino} gives a lifetime for the process $\tau
(^{12}C \rightarrow ^{12}\widetilde{C} + \gamma)> 2.1\times
10^{27}~\textrm{years}$ where $^{12}\widetilde{C}$ is a hypothetical
Pauli-forbidden nucleus with an extra nucleon in the filled
$K$-shell of $^{12}C$. This corresponds to a branching ratio of the
order of $10^{-58}$.

In the Kamiokande~\cite{SKamiokande} experiment searches were made
for forbidden transitions in $^{16}O$ nuclei and they obtain a bound
on the ratio of forbidden transitions to normal transitions. This
branching ratio is $< 2.3 \times 10^{-57}$.

The NEMO collaboration~\cite{NEMO} searches for anomalous
$^{12}\widetilde{C}$ atoms which are those with $3$ $K$-shell
electrons. The bound on the existence of such atoms is
$\frac{^{12}\widetilde{C}}{^{12}C}<2.5\times 10^{-12}$.

NEMO-2~\cite{NEMO-2} gives a lifetime $> 4.2\times 10^{24}~
\textrm{years}$ for a $^{12}C$ nuclear process which corresponds to
a branching ratio of the order $<10^{-55}$.

Atomic transition experiments have been conducted in Maryland using
copper (Cu) atoms. The idea here is to introduce new electrons into
a copper strip and to look for the K X-rays that would be emitted if
one of these electrons were to be captured by a Cu atom and cascade
down to the $1S$ state despite the fact that the $1S$ level was
already filled with two electrons. The probability for this to occur
was found to be less than $1.76\times 10^{-26}$~\cite{Ramberg}.

An improved version of the experiment at Maryland has been performed
by the VIP collaboration~\cite{VIP}. They improved the limit
obtained by Ramberg and Snow at Maryland by a factor of about $40$.
The limit on the probability of PEP violating interactions between
external electrons and copper is found to be less than $4.5\times
10^{-28}$.

Some of the above experiments give only lifetimes for the forbidden
processes. To obtain the branching ratio in such cases we multiply
the given rate with the typical lifetimes for such  processes. In
the case of an atomic process we use the number $10^{-16}$ seconds
and for a nucler process we use $10^{-23}$ seconds.

$5.2\times 10^{13}~\textrm{m}^{-1}$ was the value of $\Delta E$ for
$^{12}C$ used in calculating the bounds on $\chi$.

The nuclear experiment by the Borexino collaboration gives the best
bound for the forbidden process to date. The bound on $\chi$
obtained from this number gives a number which is much greater than
Planck energy.

\begin{table}
\begin{center}
\begin{tabular}{|c|c|c|c|}
\hline Experiment & Type & Bound on $\chi$ (Length scales) & Bound on $\chi$ (Energy scales) \\ \hline \hline Borexino & Nuclear & $\lesssim
10^{-43}~\textrm{m}$ & $\gtrsim
10^{24}~\textrm{TeV}$  \\
Kamiokande & Nuclear & $10^{-42}~\textrm{m}$ & $10^{23}~\textrm{TeV}$
\\ NEMO & Atomic & $10^{-12}~\textrm{m}$ & $10^{5}~\textrm{eV}$
\\ NEMO-2 & Nuclear & $10^{-41}~\textrm{m}$ & $10^{22}~\textrm{TeV}$
\\ Maryland & Atomic & $10^{-20}~\textrm{m}$ & $10~\textrm{TeV}$
\\ VIP & Atomic & $10^{-21}~\textrm{m}$ & $100~\textrm{TeV}$
\\
\hline
\end{tabular}
\end{center}
\caption{Bounds on the noncommutativity parameter $\chi$}
\label{table1}
\end{table}

\section{Remarks}
We conclude with the following remark. Corrections to the rate of a
generic process from spacetime noncommutativity is $O(\chi ^2E)$
where $E$ is a typical energy for the transition in question and
$\chi^2$ is of the order of the square of the Planck length. The
corresponding time scales are very long for conventional estimates
of the Planck length and this is precisely in agreement with numbers
from available experiments.

Phenomenological models of large extra dimensions (see
\cite{Shifman:2009df} for a pedagogical introduction to models of
large extra dimensions) and Randall-Sundrum scenario
\cite{Randall:1999ee} (see also \cite{Randall:2005xy}, intended for
a general audience) bring the scale of new fundamental physics from
$10^{16}$ or $10^{19}$ GeV down to $10$ or $100$ TeV scales. If the
effective four-dimensional (reduced) Planck energy scale is in the
TeV range, these time scales are very short, and may be of the order
of $10^{-18}~\textrm{sec}$. So for these processes, the earth's
movements are adiabatic (not sudden). By the adiabatic theorem, we
expect that $\tau_{\chi\vec{n}}$ eigenstates will smoothly evolve
into $\tau_{\chi\vec{m}}$ eigenstates of the same eigenvalue. No
non-Pauli effects can thus be hoped for.

Lifetimes for non-Pauli transitions, which create Pauli-forbidden
levels, are much longer than the age of the universe in our model.
So if there were only Pauli-allowed levels at the initiation of the
universe, there will not be a significant amount of non-Pauli levels
now. Hence no conflict with experiment from the lack of abundance of
these levels is expected.

\section{Acknowledgements}
We are very grateful to Gianpiero Mangano, who helped us with
important suggestions at every stage of this work. This work has
taken this shape only because of his critical comments.

A.P.B. and P.P. thank Prof. T.R. Govindarajan for the hospitality at
IMSc, Chennai where this work was completed. This work was supported
in part by the U.S. Department of Energy grant under the contract
number DE-FG02-85ER40231. The work of A.P.B. was also supported by
the Department of Science and Technology, India.

\end{document}